# Enhancement in Quality Factor of SRF Niobium Cavities by Material Diffusion

Pashupati Dhakal, Gianluigi Ciovati, Peter Kneisel and Ganapati Rao Myneni, Jefferson Lab, Newport News VA

*Abstract*— An increase in the quality factor of superconducting radiofrequency cavities is achieved by minimizing the surface resistance during processing steps. The surface resistance is the sum of temperature independent residual resistance and temperature/material dependent Bardeen-Cooper-Schrieffer (BCS) resistance. High temperature heat treatment usually reduces the impurities concentration from the bulk niobium, lowering the residual resistance. The BCS part can be reduced by selectively doping non-magnetic impurities. The increase in quality factor, termed as *Q*-rise, was observed in cavities when titanium or nitrogen thermally diffused in the inner cavity surface.

## I. INTRODUCTION

Superconducting radiofrequency (SRF) technology is being used not only for the basic physics research but also for applications that have benefited society. Future accelerating machine heavily rely on the SRF technology.

Few examples of the current and future projects based on SRF technology are continuous wave (CW) and pulsed free electron lasers, x-ray laser oscillators, energy recovery linac (ERL) based light sources, short photon pulses in storage ring light sources, electrons and ions colliders, accelerator driven systems (ADS) for medical isotope production and nuclear waste transmutation. SRF technology is currently based on niobium superconducting hollow resonating structures ("cavities") to accelerate the beam of charged particles.

The superiority of superconducting cavities over those made of normal-conducting metal is its ability to store large amount of energy with much lower dissipation. The performance of SRF cavities are measured in terms of the quality factor $Q_0=G/R_s$, where $G$ is the geometric factor which depends on the cavity geometry and $R_s$ is the average surface resistance of the inner cavity wall, as a function of the accelerating gradient, $E_{acc}$. For typical Nb cavities resonating at frequencies 0.5–2 GHz operating at a temperature of ~2.0 K, the quality factor is in the $10^{10}$–$10^{11}$ range. In order to meet ever more challenging requirements of current and future accelerators, improvements of quality factor and accelerating gradient are the main topics of research on SRF cavities.

Surface treatments of bulk Nb SRF cavities include several cycles of mechanical, thermal and chemical processing steps [1, 2]. These preparation steps are crucial to mitigate phenomena such as field emission, multipacting, *Q*-slope and premature quench. One of those treatments is a high temperature (600-800 °C) annealing in high vacuum to degas the hydrogen from the bulk. Results on SRF cavities which were heat treated in the temperature range of 600–1600 °C without following chemical etching, showed improvements in the quality factor compared to standard preparation methods [3, 4], by minimizing both the residual and BCS surface resistance. It was recently discovered that, introducing interstitials such as titanium [5] and nitrogen [6], which can trap hydrogen [7,8], in the Nb surface seems to be beneficial to achieve higher quality factor. These trapping centers not only minimize the mobile hydrogen in niobium but also reduce the residual resistivity ratio (RRR) of inner cavity surface and hence the reduction in BCS surface resistance. A rise in the quality factor as a function of accelerating gradient has also been observed in such "doped" Nb cavities, in contrast to a degradation of $Q_0$ with field which is typically measured after standard preparation. The analysis of the temperature and field dependence of the surface resistance of a titanium doped cavity showed that the *Q*-rise is consistent with broadening of the peaks at the gap edges in the electronic density of states of "dirty" Nb by the rf current [9]. In this contribution, we present the results of studies of material diffusion in SRF cavities made of Nb with different bulk purity and grain structure.

## II. EXPERIMENTAL RESULTS

Several niobium cavities of frequency 1.3 and 1.5 GHz made from ingot (large grain) Nb with different residual resistivity (RRR) values and fine-grain reactor grade niobium were used for this study. The cavities were heat treated in an induction furnace [10] up to 1400 °C as well as in a standard radiative heating production furnace at 800 °C in the presence of titanium[11] and nitrogen to explore the effect of the material diffusion on quality factor of niobium cavities. Several niobium samples (5 × 7 × 3 mm$^3$) were also heat treated along with the cavities to determine the depth of material diffusion.

*A. Titanium Diffusion*

Historically titanium has been used to post-purify niobium SRF cavities during heat treatment since it is a good solid-state getter for most impurities dissolved in Nb. After the titanification of SRF niobium, the inner surface of the cavity is chemically etched to remove the impurities gettered by evaporated titanium. This process is beneficial for the increase in RRR of Nb and therefore in better thermal stability of SRF cavities. We have observed that smaller concentration of titanium (~1 at.%) diffused into the inner surface of a cavity resulted in the quality factor increasing with field up to $B_p$~90mT [4]. In this investigation, the titanium was sublimated from the cavity flanges made of Ti45Nb alloy. By this process the $Q_0$-value was improved by 2–4 times compared to the baseline chemically etched cavities as shown in Fig. 1.

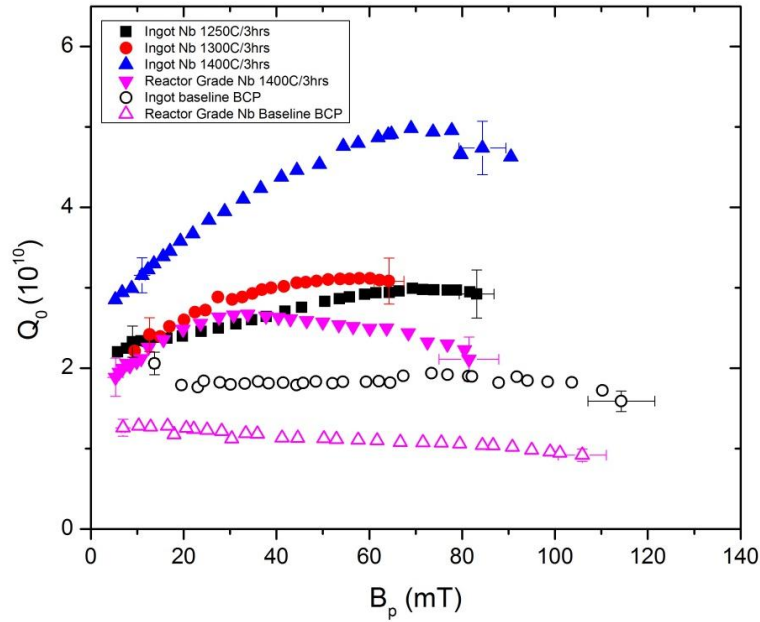

Fig. 1. $Q_0$(2 K) vs $B_p$ for the ingot and reactor grade Nb 1.5 GHz ($B_p/E_{acc}$ = 4.43 mT/(MV/m)) SRF cavities heat treated in titaniumenvironment in temperature range of 1250–1400 °C. These rf tests were limited by cavity quench. About 20 μm inner surface of ingot cavity was removed by BCP before each heat treatment.

Several cavities with Ti45Nb flanges made from RRR~100-200 ingot niobium were also heat treated at different temperatures to reproduce the $Q$-rise phenomenon [11]. Future work will focus on a better control of the doping process by introducing a pure titanium rod inside single-cell cavities during heat treatment.

*B. Nitrogen Diffusion*

Thermal diffusion of nitrogen into Nb is a method which had been used in the past to produce NbN films [12,13,14,15], with higher superconducting transition temperature than Nb. This method was recently pursued at Fermi Lab by heat treating 1.3 GHz single-cell cavities at 800-1000°C in 20-50 mTorr nitrogen pressure but resulted in $Q_0$-values of $10^7$-$10^8$ at 2.0 K. However, it was discovered that subsequent material removal by electropolishing (EP) led to a $Q$-rise as high as ~$7\times10^{10}$ at 2 K with increasing rf field up to ~20 MV/m, even though no NbN stoichiometric phase was found [6]. A collaboration between Jefferson Lab, Fermi Lab and Cornell University began to investigate the robustness of this process to meet the specifications for cavities for the LCLS-II project, requiring a $Q_0$ value of at least $2.7\times10^{10}$ at 2.0 K and a gradient of $E_{acc}$=16.0 MV/m in 9-cell, 1.3 GHz cavities [16]. The work of this collaboration is focused on developing the process on fine-grain (ASTM > 5), high purity (RRR>300) Nb cavities.

Following the development of the recipe of nitrogen doping at Jefferson Lab [17] for the LCLS-II project, we explored whether a similar process could be applied to:

1. cavities made of reactor-grade Nb, and
2. cavities made of ingot Nb of different purity.

Ingot Nb is an alternative material for the fabrication of SRF cavities having grain size of few cm$^2$ [18]. The use of bulk Nb with lower purity than standard RRR > 300 would result in significant material cost savings. Two single-cell 1.5 GHz cavities, one made of fine-grain, reactor grade (RRR~100) Nb from Cabot and one made of medium purity (RRR~200) ingot Nb from CBMM were heat treated at 800 °C for 3 hours and at the end of the heat treatment nitrogen partial pressure of ~20 mTorr was injected in the furnace for 2 min. After 2 min, the furnace is evacuated and further annealed at 800 °C for 10 minutes continues to diffuse the absorbed nitrogen. Both cavities were electropolished at the same time to remove ~7 μm from the inner surface, followed by high pressure rinse with ultra-pure water and cryogenic RF testing. The summary of the RF test results is shown in Fig. 2.

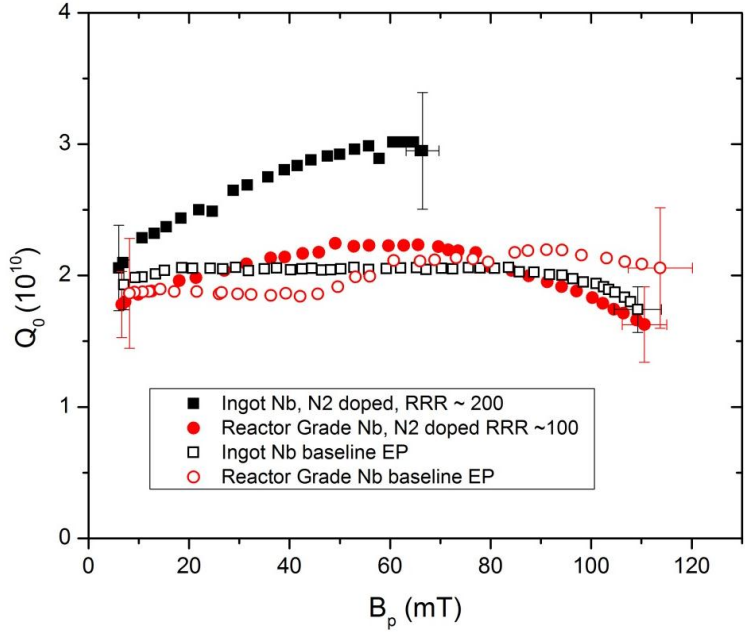

Fig. 2. $Q_0$(2 K) vs $B_p$ for the ingot and reactor grade Nb 1.5 GHz ($B_p/E_{acc}$ = 4.43 mT/(MV/m)) SRF cavities heat treated in the presence of nitrogen gas followed by 7 μm EP. These rf tests were limited by cavity quench.

We have also heat treated two single-cell 1.3 GHz cavities made from ingot niobium with RRR> 300 from Tokyo-Denkai. The cavities were heat treated at 800 °C for 3 hours followed by 20 minutes of exposure to nitrogen at pressure of ~25 mTorr. The nitrogen is then evacuated and the cavities were further annealed at 800 °C for 30 min. Cavities' inner surface was electropolished to remove ~10 μm, followed by high pressure rinse with ultra-pure water. The summary of the RF test results before and after the nitrogen doping and EP are shown in Fig. 3. The $Q$-rise was observed after the nitrogen doping and EP as opposed to the $Q$-degradation measured in the baseline measurements. However, the quench field is degraded by ~40% compared to the baseline. Repeatedly quenching the cavities more than a hundred times, resulted in an increase of the residual resistance of only ~0.5 nΩ. The results shown in Fig. 3 are very similar to those obtained on fine-grain, RRR>300 cavities [6, 16, 17].

We began exploring the possibility of streamlining the doping procedure by eliminating the EP removal. This requires finding the optimum time, temperature and pressure parameters to produce a nitrogen diffusion profile in Nb which would result in a $Q$-rise. This study was done on three 1.5 GHz single-cell cavities, one (labelled "P2") made of high-purity, fine-grain Nb, the other two made of medium-purity (RRR~200) ingot Nb from CBMM and labelled "LG-B" and "SC-IB". The following parameter space was explored: 800-1250 °C/2-3h heat-treatment temperature/time, $10^{-8}$-$10^{-3}$ Torr/10 min – 3h nitrogen pressure/time, 800 °C/10-30 min diffusion temperature/time. After nitrogen doping, the cavities were just degreased and high-pressure rinsed with DI water. The results from rf tests at 2.0 K showed either no change from the baseline performance or a degradation of $Q_0(B_p)$, as shown for example in Fig. 4.

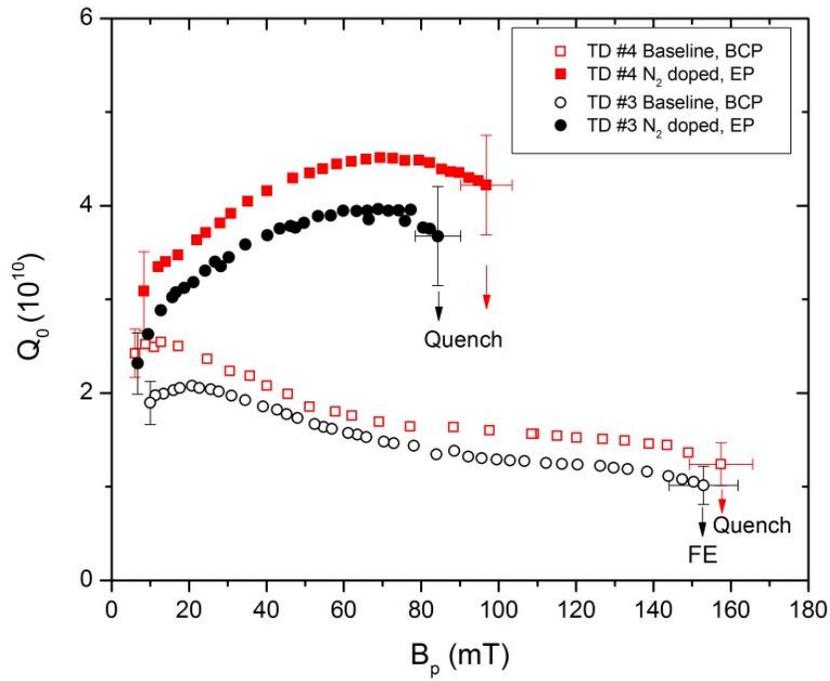

Fig. 3. $Q_0$(2 K) vs $B_p$ for ingot Nb with RRR>300 1.3 GHz ($B_p/E_{acc}$ = 4.33 mT/(MV/m)) cavities heat treated in the presence of nitrogen gas followed by ~10 μm EP.

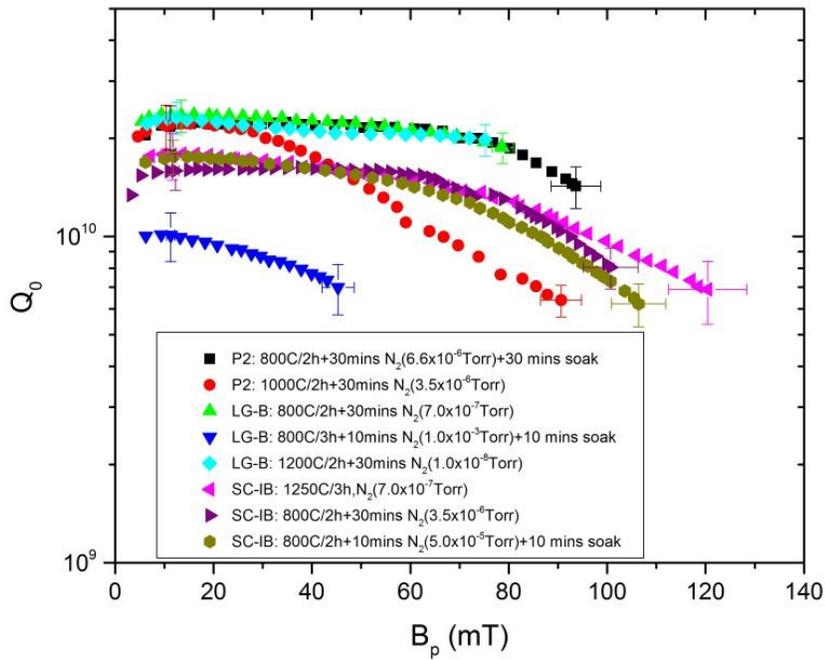

Fig. 4. $Q_0$(2 K) vs $B_p$ measured on three different 1.5 GHz single-cell cavities after nitrogen doping without post-EP with the doping parameters listed in the legend.

## C. Samples Surface Analysis

Several samples have been heat treated with the cavities to study the depth profile and concentration of the material diffused in the niobium surface using secondary-ion mass spectrometry (SIMS) for both titanium and nitrogen. Titanium depth profiles have been published already in [4].

Figure 5 shows an example of NbN$^-$/Nb$^-$ depth profiles on samples heat treated with nitrogen without post-EP. The sample heat treated with 20 mTorr nitrogen was processed with cavities which were subjected to 7 μm post-EP and additional SIMS measurements will be made to extend the depth profile to at least 7 μm.

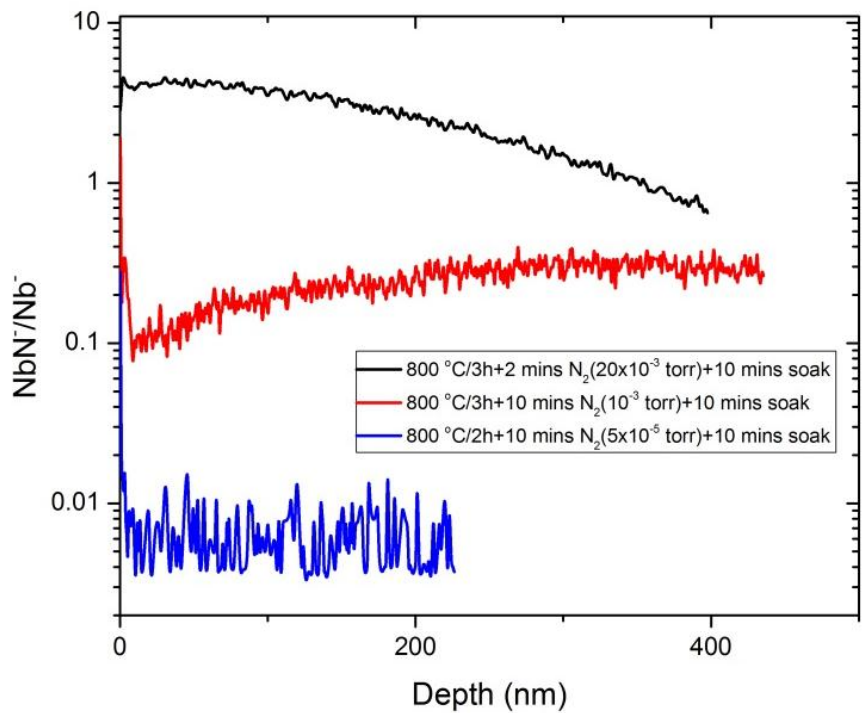

Fig. 5. The ratio of NbN$^-$/Nb$^-$ counts as a function of depth profile measured by Times-of-Flight SIMS in samples heat treated at 800 ºC with different partial pressure of nitrogen.

## III. DISCUSSION

The results presented in Sec. II show the improvement of $Q_0$-value compared to that obtained after standard treatments by doping bulk ingot Nb cavities of medium to high RRR, with titanium, without any subsequent chemical etching, or by N-doping followed by EP. Both processes resulted in $Q_0$-values increasing with increasing rf fields up to ~20 MV/m. The $Q_0$-improvement on reactor-grade cavities was not as pronounced as in ingot cavities suggesting that the doping process parameters would have to be modified for such material, since this material already has higher concentration of impurities. As discussed in ref. [11], a better control of the titanium doping process has to be used in order to reproducibly obtain $Q_0$-values of ~3×10$^{10}$ at 2.0 K and ~20 MV/m. Our preliminary parameter study on nitrogen doping without subsequent EP hasn't been successful in improving the quality factor. Further study on sample analysis is needed to determine the amount of nitrogen introduced into the cavity surface and compare it to that obtained by doping at higher nitrogen pressure followed by EP.

High temperature heat treatments of cavities without subsequent chemical etching had also shown ~20% improvement in the quality factor compared to that measured after standard treatments and samples' analysis showed a correlation with reduced hydrogen concentration [3, 4]. Precipitation of lossy NbH can still be a contributor to the residual resistance, even after degassing at 800 °C because of subsequent chemical etching. It has been known that titanium or nitrogen doping on Nb effectively traps mobile hydrogen and prevents precipitation of hydrides [7, 8]. In addition, diffusion of impurities (nitrogen, and titanium) increases the residual dc resistivity (reduction in RRR) of the Nb and changes the superconducting behavior towards the so-called "dirty limit" (electronic mean free path much less than the coherence length, $l \ll \xi$). Curve fits of $Q_0(T)$ data indicated a reduction of mean free path after the interstitial diffusion. The most outstanding feature in the rf test results after doping is the extended $Q_0$-rise.

Even though there is an increase in quality factor, the maximum gradient of cavities doped by titanium or nitrogen is often limited to the much lower values compare to those achieved by standard treatments. This is probably due to the early vortex penetration due to the reduction of the lower critical field by doping. Nitrogen doped fine grain cavities tend to trap higher ambient magnetic flux during the cool down process resulting in a higher residual resistance [19]; this suggests that a better magnetic shielding environment is needed to maintain the high quality factor during operation. Further investigation to engineer SRF niobium via material diffusion is in progress in attempts to push the gradient > 30 MV/m while preserving the high quality factor.

## IV. CONCLUSION

Significant improvement in the quality factor has been achieved in medium and high purity ingot Nb cavities by material (titanium and nitrogen) diffusion during the high temperature heat treatment. This improvement can be explained by a reduction of the residual resistance by preventing the precipitation of hydrides and by a reduction of the BCS surface resistance by lowering the electronic mean free path. Further material characterization is planned to understand and optimize the doping process.

**ACKNOWLEDGEMENT**

We would like to acknowledge J. Folkie, P. Kushnick, D. Forehand, B. Clemens, G. Slack, B. Martin and T. Harris at Jefferson Lab for helping with EP, HPR, HT, cryogenic test and cavity fabrication. We would like to thank C. Reece and A. Palczewski for discussion about the nitrogen doping recipe. We would also like to thank E. Zhou of the Analytical Instrumentation Facility at NCSU for the TOF-SIMS measurements.